\documentstyle[12pt,a4wide]{article}
\newcommand{\be}{\begin{equation}}
\newcommand{\ee}{\end{equation}}
\newcommand{\bea}{\begin{eqnarray}}
\newcommand{\eea}{\end{eqnarray}}
\newcommand{\bean}{\begin{eqnarray*}}

\newcommand{\eean}{\end{eqnarray*}}
\font\upright=cmu10 scaled\magstep1 \font\sans=cmss10
\newcommand{\ssf}{\sans}
\newcommand{\stroke}{\vrule height8pt width0.4pt depth-0.1pt}
\newcommand{\Z}{\hbox{\upright\rlap{\ssf Z}\kern 2.7pt {\ssf Z}}}

\newcommand{\C}{{\rlap{\rlap{C}\kern 3.8pt\stroke}\phantom{C}}}
\newcommand{\R}{\hbox{\upright\rlap{I}\kern 1.7pt R}}
\newcommand{\CP}{\C{\upright\rlap{I}\kern 1.5pt P}}
\newcommand{\PP}{\hbox{\upright\rlap{I}\kern 1.5pt P}}

\newcommand{\identity}{{\upright\rlap{1}\kern 2.0pt 1}}

\newcommand{\HH}{\mbox{\hbox{\upright\rlap{I}\kern 1.7pt H}}}

\newcommand{\fr}{\frac}

\newcommand{\ra}{\rightarrow}
\newcommand{\al}{\alpha}
\newcommand{\bt}{\beta}
\newcommand{\pr}{\partial}

\newcommand{\vphi}{{\varphi}}

\input{epsf}
\begin{document}

\title{\bf Spinning Gravitating Skyrmions}
\vspace{1.5truecm}
\author{
{\bf Theodora Ioannidou$^a${\footnote{{\it Email}: ti3@auth.gr}},
 \bf Burkhard Kleihaus$^b${\footnote{{\it Email}:
 kleihaus@theorie.physik.uni-oldenburg.de}}
and Jutta Kunz$^b${\footnote{{\it Email}:
 kunz@theorie.physik.uni-oldenburg.de}}
}\\
$^a$ Mathematics Division, School of Technology\\
Aristotle University of Thessaloniki \\
Thessaloniki 54124,  Greece\\
$^b$ Institut f\"ur  Physik, Universit\"at Oldenburg, Postfach 2503\\
D-26111 Oldenburg, Germany
}

\vspace{1.5truecm}

\date{\today}

\maketitle

\vspace{1.0truecm}

\begin{abstract}
We investigate self-gravitating rotating solutions in the
Einstein-Skyrme theory.
These solutions are globally regular and
asymptotically flat. We present a new kind of solutions with zero
baryon number, which possess neither a flat limit nor a static limit.
\end{abstract}


\section{Introduction}

In non-Abelian field theories coupled to gravity particle-like solutions
as well as black holes arise \cite{overview}. The latter are of importance
as counterexamples to the no-hair conjecture.
In recent years, in particular, various stationary rotating non-Abelian black holes
have been  studied \cite{rotbh}.
However, the construction of stationary rotating particle-like solutions is
still a difficult task.
Although existence of such solutions in non-Abelian gauge field theories
has been restricted \cite{vdBijRadu,VolWoh,vdBijRadu2},
they can exist in the topologically trivial sector \cite{PRT,KKN}.
On the other hand, stationary rotating soliton solutions in flat space
have been obtained in the Skyrme model \cite{BKS} and the
U(1) gauged  Skyrme model \cite{RaTch} in the nontrivial sector.

The Skyrme model is a nonlinear chiral field theory in which
baryons and nuclei are described in terms of solitons (so-called Skyrmions).
Due to  the long-standing difficulties in finding a
satisfactory theoretical model for the interaction of baryons,
much effort has been devoted to the study of classical and
quantised interactions of  Skyrmions.
However, the quantization of the Skyrme model is not only difficult since it is
a non-renormalizable field theory; but also spinning Skyrmions
 must be considered which means that the solutions must consist of massive pions.
This follows from the fact that the Skyrmion can only spin at
a frequency up-to the pion mass before it begins to radiate pions \cite{spinSk}.
Recently in \cite{BKS}, it was shown numerically that a good description of
protons and neutrons can be achieved with spinning Skyrmions, provided the pion mass
is chosen twice the experimental value.

The rotating Skyrmion solutions of \cite{BKS} are expected to persist, when
the coupling to gravity is turned on gradually, analogous to the static
Skyrmion solutions \cite{gravSk}.
In the static limit,
a branch of gravitating Skyrmions emerges from the
flat space Skyrmion, when the coupling to gravity is increased from zero
\cite{gravSk}. This branch terminates at a maximal value of the
coupling parameter, when the coupling to gravity becomes too large
for solutions to persist. A second branch of solutions exists which  merges with
the first one at the maximal value of the coupling parameter and extends
back to zero. The solutions on the second branch possess a larger mass
and they are unstable \cite{Heussler}.
In the limit of vanishing coupling
the solutions shrink to zero size and their mass diverges.
As shown in \cite{Bizon}, in this limit the Skyrmion solutions approach the
 lowest mass Bartnik-McKinnon (BM) solution of the  $SU(2)$ Einstein-Yang-Mills
 theory \cite{BM}.

In this paper we investigate the stationary rotating generalisation of the
static gravitating Skyrmions. We show that in contrast to the static case,
additional branches of solutions arise, which are not related to the
flat space Skyrmion or to the BM solution.
Most interestingly, we find a new kind of solution with zero baryon number,
which exists for arbitrary (finite) coupling parameter, but does
not possess a flat space limit.
In particular, Section 2 presents the Einstein-Skyrme Lagrangian and the
ansatz for the Skryme field and the metric, which lead to
stationary rotating Skyrmions. In Section 3 the numerical solutions
are discussed, while the conclusions are given in Section 4.

\section{Einstein-Skyrme Theory}

The $SU(2)$ Einstein-Skyrme Lagrangian reads
 \be
  {\cal L}=
  \fr{R}{16\pi G}
 +\fr{\kappa^2}{4} \mbox{Tr}\left(K_{\mu}\,K^\mu \right)
 +\fr{1}{32e^2}\mbox{Tr}\left(
     \left[K_\mu,K_\nu\right]\left[K^\mu,K^\nu\right]
                        \right)
 +\fr{m_\pi^2}{2}\mbox{Tr}\left(\frac{U+U^\dagger}{2}-\identity\right),
  \label{lag}
  \ee
and its action is given by
 \be
  S=\int {\cal L}\sqrt{-g}\, d^4x.
  \label{ac}
  \ee
Here $R$ is the curvature scalar, $G$ is the Newton constant,
$\kappa$ and $e$ are the Skyrme model coupling constants,
$m_\pi$ is the pion mass, and $g$ corresponds to
the determinant of the metric.
The $SU(2)$ Skyrme field $U$ enters via
 $K_\mu=\pr_\mu U U^{-1}$.

Variation of (\ref{ac}) with respect to the metric
$g^{\mu \nu}$ leads to the Einstein equations
\bea
G_{\mu \nu}&=&R_{\mu \nu}-\fr{1}{2}g_{\mu \nu}R\\\nonumber
&=&8\pi G T_{\mu \nu},
\label{E}
\eea
where the stress-energy tensor is given by
\begin{eqnarray}
T_{\mu\nu}\!\!\!\! & = &\!\!\!\!
-\fr{\kappa^2}{2}
\mbox{Tr}\!\left(K_\mu K_\nu-\fr{1}{2}g_{\mu \nu} K_\al K^\al \right)
\!-\! \fr{1}{8e^2}
\mbox{Tr}\!\left(
g^{\al \bt}\left[K_\mu,K_\al\right]\left[K_\nu,K_\bt\right]\!-\!
\fr{1}{4}g_{\mu \nu} \,\left[K_\al,K_\bt\right]\left[K^\al,K^\bt\right]
\right)
\nonumber\\
 & &
+ g_{\mu \nu}m_\pi^2\mbox{Tr}\left(\frac{U+U^\dagger}{2}- \identity\right).
\label{T}
\end{eqnarray}

For stationary rotating solutions, two commuting Killing vector
fields are imposed on the space-time: $\xi = \partial_t$
and $\eta = \partial_\varphi$ in a system of adapted coordinates
$(t, r, \theta, \varphi)$. In these coordinates
the metric can be expressed in Lewis-Papapetrou form
\begin{equation}
ds^2=-fdt^2+\fr{m}{f}
\left(dr^2+r^2 d\theta^2\right)
+l r^2 \sin^2\theta \left(d\vphi-\frac{\omega}{r}dt\right)^2,
 \label{s}
\end{equation}
where $f$, $m$, $l$ and $\omega$ are functions of $r$ and $\theta$ only.

Then, the total mass and angular momentum are defined by
\begin{equation}
{\cal M} = \frac{1}{4 \pi G} \int_\Sigma R_{\mu\nu}\, k^\mu\, \xi^\nu dV ,
 \ \ \ \
{\cal J} = -\frac{1}{8 \pi G} \int_\Sigma \,R_{\mu\nu} \,k^\mu \eta^\nu dV ,
\label{massJ}
\end{equation}
respectively. Here $\Sigma$ denotes an asymptotically flat hyper-surface,
$dV$ is the natural volume element on $\Sigma$, $k^\mu$ is normal to $\Sigma$
and $k_\mu k^\mu = -1$.

In order for finite energy configurations to exist the Skyrme field
must tend to a constant matrix at spatial infinity:
$U \ra \identity$ as
$|x^\mu| \ra \infty$.
This effectively compactifies the three-dimensional Euclidean space
into $S^3$ and implies that the Skyrme fields
can be considered as maps from $S^3$ into $SU(2)$.
As the third homotopy class of $SU(N)$ is $Z$, every field configuration
is characterized by a topologically invariant integer $B$,
which can be obtained as
\begin{equation}
B = \int_{\Sigma} B^\mu k_\mu dV  ,
\label{B}
\end{equation}
where $B^\mu$ is the topological current
\begin{equation}
B^\mu = \frac{1}{\sqrt{-g}} \frac{1}{24\pi^2}
\epsilon^{\mu \nu \alpha \beta} {\rm Tr}\,
\left( K_\nu K_\alpha K_\beta \right) . \end{equation}
This winding number classifies the solitonic
sectors in the model and may be identified with the  baryon number of the field
configuration.

For spinning Skyrmion the ansatz  is of the form\footnote{
Strictly speaking, the ansatz is neither stationary nor axially
symmteric, since it depends explicitly on time and the azimuthal angle.
However, the stress-energy tensor does possess the
corresponding symmetries. See also Ref. \cite{VolWoh}.}:

\begin{equation}
U = n_1 \identity + i n_3 \tau_z
    + i n_2 \left(\tau_x \cos(\vphi + \omega_s t) + \tau_y
                   \sin(\vphi + \omega_s t)  \right),
\label{Umat}
\end{equation}
where $\tau_x$ ,$\tau_y$, $\tau_z$ are the Pauli matrices; $n_i$
are functions of $r$ and $\theta$ only, satisfying the constraint
$ C:=\left(1-\sum n_i^2\right) = 0$,
and  the constant $\omega_s$ corresponds to the spinning frequency of
the Skyrmions.

When deriving the partial differential equations (PDEs) for the
Skyrmion functions $n_i$, we have to take into account the constraint
$C = 0$.
This can be achieved by  adding the constraint multiplied by some constant,
say $c_0$, to the Lagrangian and deriving the variational equations:
\begin{equation}
E_i  = \partial_r\frac{\partial {\cal L}\sqrt{-g}}{\partial(\partial_r n_i)}
      +\partial_\theta\frac{\partial {\cal L}\sqrt{-g}}{\partial
(\partial_\theta n_i)}-\frac{\partial {\cal L}\sqrt{-g}}{\partial n_i}
      +2 c_0 n_i \sqrt{-g} =0 \ .
 \label{PDEs}
\end{equation}
Then,  the constant $c_0$ can be obtained from the linear superposition
$ \sum n_i E_i = 0$, and substituted back in the PDEs of (\ref{PDEs}).

\section{Numerical Solutions}

\subsection{Parameters and Boundary Conditions}

Introducing the dimensionless radial coordinate $x=\kappa e r$,
the gravitational coupling parameter $\alpha^2 = 4 \pi G\kappa^2$,
the spinning frequency $\hat{\omega}_s = \omega_s/ \kappa e$,
and the pion mass $\hat{m}_\pi = m_\pi/ \sqrt{\kappa/e}$,
action (\ref{ac}) becomes
\be
  S=\fr{\kappa}{e}
  \int \left[
  \fr{R}{4\alpha^2}
 +\fr{1}{4} \mbox{Tr}\left(K_{\mu}\,K^\mu \right)
 +\fr{1}{32}\mbox{Tr}\left(
     \left[K_\mu,K_\nu\right]\left[K^\mu,K^\nu\right]
                        \right)
 +\fr{\hat{m}_\pi^2}{2}\mbox{Tr}\left(\frac{U+U^\dagger}{2}-\identity\right)
    \right]
                \sqrt{-g}\, d^4x \ ,
  \label{ac_n}
\ee
while the Einstein equations read: $G_{\mu\nu} = 2 \alpha^2 T_{\mu\nu}$.
We also introduce the dimensionless mass
$M ={\cal M}/[4\pi \kappa/e]$ and angular momentum
$J= {\cal J}/[4\pi/e^2]$.
That way, the solutions depend only on the parameters  $\alpha$, $\hat{\omega}_s$
and $\hat{m}_\pi$. For convenience we will rename $\hat{\omega}_s\to \omega_s$.

At the origin, the boundary conditions are
\be
n_1(0) = -1  , \ \ n_2(0) = n_3(0) = 0  ,
\ \ \partial_x f|_0 = 0  , \ \ \partial_x l|_0 = 0   ,
\ \ \partial_x m|_0 = 0  , \ \ \omega(0) = 0   ,
\ee
while for large $x$, since the asymptotic value of the  Skyrme field is the
 unit matrix and of the metric is the Minkowski metric, we get
\be
n_1(\infty) \to 1, \ \ n_i(\infty)|_{i=2,3} \to 0 ,
\ \ f(\infty) \to  1  , \ \  l(\infty)\to 1  ,
\ \ m(\infty) \to  1   , \ \  \omega(\infty)\to 0.
\ee
On the $z$-axis ($\theta=0$)  the boundary conditions follow from regularity
\begin{eqnarray}
& &
\partial_\theta n_1|_{\theta=0} = 0  ,  \ \
n_2(\theta=0) = 0  , \ \
\partial_\theta n_3|_{\theta=0} = 0  ,
\nonumber \\
& &
\partial_\theta f|_{\theta=0} = 0  , \ \
\partial_\theta l|_{\theta=0} = 0  , \ \
\partial_\theta m|_{\theta=0} = 0  , \ \
\partial_\theta \omega|_{\theta=0} = 0  ,
\label{bc-zaxis}
\eea
while in the $xy$-plane ($\theta=\pi/2$) from  reflection symmetry
\bea
& &
\partial_\theta n_1|_{\theta=\pi/2} = 0  , \ \
\partial_\theta n_2|_{\theta=\pi/2} = 0  , \ \
n_3(\theta=\pi/2) = 0  ,
\nonumber \\
& &
\partial_\theta f|_{\theta=\pi/2} = 0  , \ \
\partial_\theta l|_{\theta=\pi/2} = 0  , \ \
\partial_\theta m|_{\theta=\pi/2} = 0  ,  \ \
\partial_\theta \omega|_{\theta=\pi/2} = 0  .
\label{bc-rhoaxis}
\end{eqnarray}

In what follows we will encounter two special cases,
the Bartnik-McKinnon solution and the non-trivial solutions
in the vacuum sector.

The first, is obtained after re-scaling $x=\alpha \tilde{x}$
and taking the limit of vanishing $\alpha$.
In this limit, the solutions are equivalent to the Bartnik-McKinnon
one with lowest mass:
\begin{equation}
n_1 = -w(\tilde{x})  , \ \
n_2 = \sqrt{1-n_1^2} \sin\theta   , \ \
n_3 = \sqrt{1-n_1^2} \cos\theta  , \ \
l=m  , \ \  \omega=0  ,
\label{EYMsol}
\end{equation}
where the gauge potential of the SU(2)
Einstein-Yang-Mills theory is parametrized from $w(\tilde{x})$  via the
relation  $A_i^a =
(1-w(\tilde{x}))\epsilon_{iaj} \tilde{x}_j/(2\tilde{x}^2)$.

The second is obtained by setting
\begin{equation}
n_1 = \cos(h)  , \ \ n_2 = \sin(h) , \ \ n_3 = 0  ,
\label{vacsol}
\end{equation}
where the function $h$ depends on $x$ and $\theta$.
Regularity and finite energy of the solutions require that $h$
 vanishes on the $z$-axis and  at infinity.
We will refer to these solutions as `pion cloud'.

\subsection{Numerical Results}

The solutions are constructed using the  software package
CADSOL \cite{fidisol} based on the Newton-Raphson algorithm.
In order to map the infinite range of the radial variable $x$
to the finite  interval $[0,1]$ we introduce the compactified radial variable
$\bar{x} = x/(1+x)$.
Typical grids contain $70\times 50$ points.
The estimated relative errors are approximately $\approx 0.1$\%,
except close to $\alpha_{\rm max}$ where they become as large as
$1$\%.

In particular, gravitating Skyrmions are constructed
and  their dependence on the coupling parameter $\alpha$
and the spinning frequency $\omega_s$ are studied
 for fixed  pion mass: $\hat{m}_\pi = 1$.
This sets a limit to the range of the spinning frequency
$\omega_s\le 1$.
We  start the discussion by a qualitative description of the
dependence of the solutions on the gravitational parameter $\alpha$
 for fixed spinning frequency $\omega_s$.
Different branches of solutions exist which are  characterized
by their limit as $\alpha$ tends to zero.
First, there are branches of solutions which tend to the flat space
Skyrmions and to the scaled BM solution which we call (for obvious reasons)
`Skyrmion' and `BM' branches, respectively.

Second, branches of solutions exist which form a  `pion cloud' for
large $x$ as  $\alpha \to 0$. In this limit the
solutions resemble a superposition of the `pion cloud' with either
a Skyrmion or the scaled BM solution and are called
 `cloudy Skyrmion' and `cloudy BM'
branches, respectively.
However, that for the  `cloudy'
solutions the limit $\alpha \to 0$ is singular, as explained below.

For fixed  $\omega_s$ the branches exist up to a maximal
value of $\alpha$, where they merge with a branch of different type.
In particular, the way  different branches merge depends on the value of
$\omega_s$ relative to the critical value $\omega_s^{\rm cr} \approx 0.9607$.
So, for $\omega_s<\omega_s^{\rm cr}$ the `Skyrmion' branches merge
 with the `BM' ones; however
for $\omega_s>\omega_s^{\rm cr}$,
the `Skyrmion'  and the `BM' branches merge with the `cloudy Skyrmion'
and the  `cloudy BM' ones, respectively.
Moreover, the `cloudy Skyrmion' branches merge with the `cloudy BM'
branches only when $\omega_s<\omega_s^{\rm cr}$.

Next a quantitative description in terms of
the dimensionless mass $M$ and the value of the
function $l_0=l(0)$ is presented.
Since with vanishing $\alpha$ the mass diverges on the `BM',
the `cloudy Skyrmions' and the `cloudy BM' branches we also consider
the scaled masses $M\alpha$ and $M\alpha^3$.

\begin{figure}[h]
\centering
\epsfysize=5.0cm
(a) \mbox{\epsffile{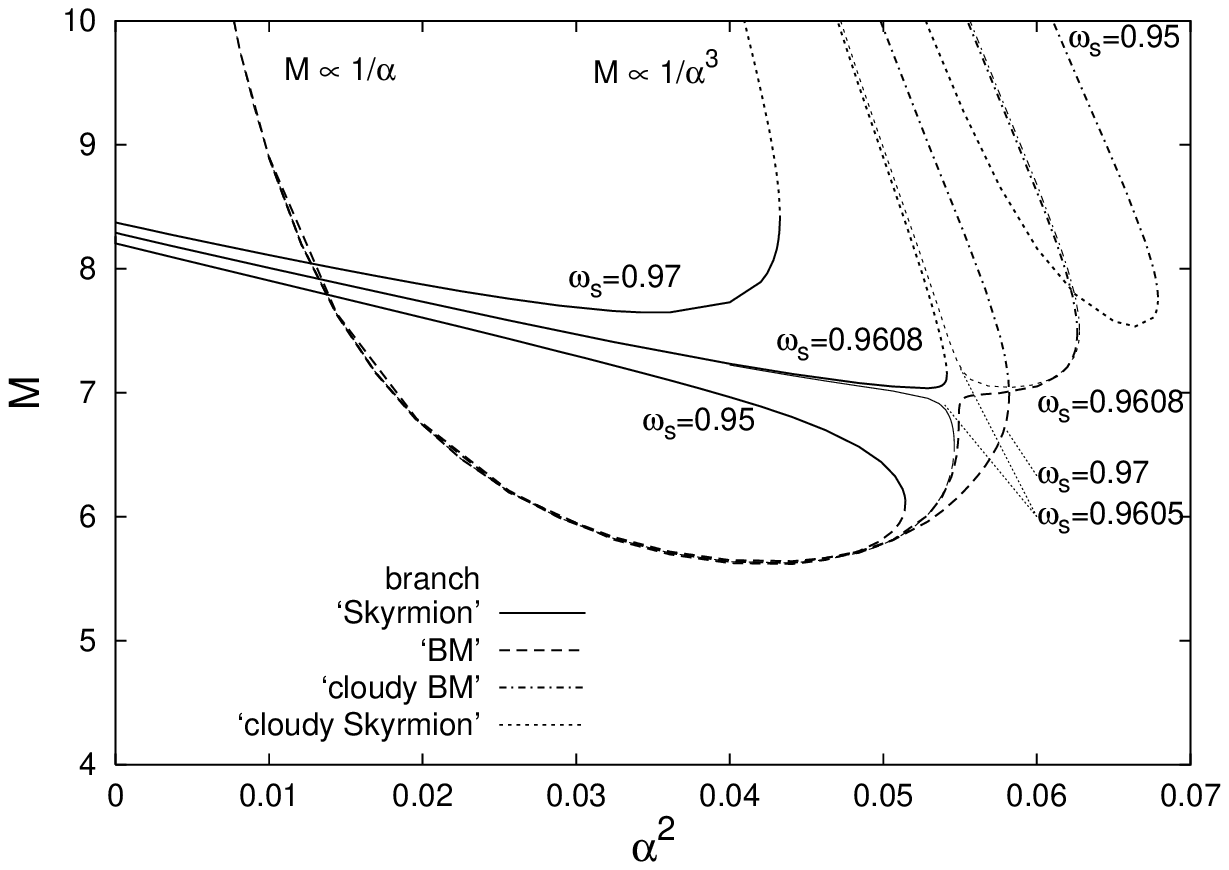}}
\epsfysize=5.0cm
(b) \mbox{\epsffile{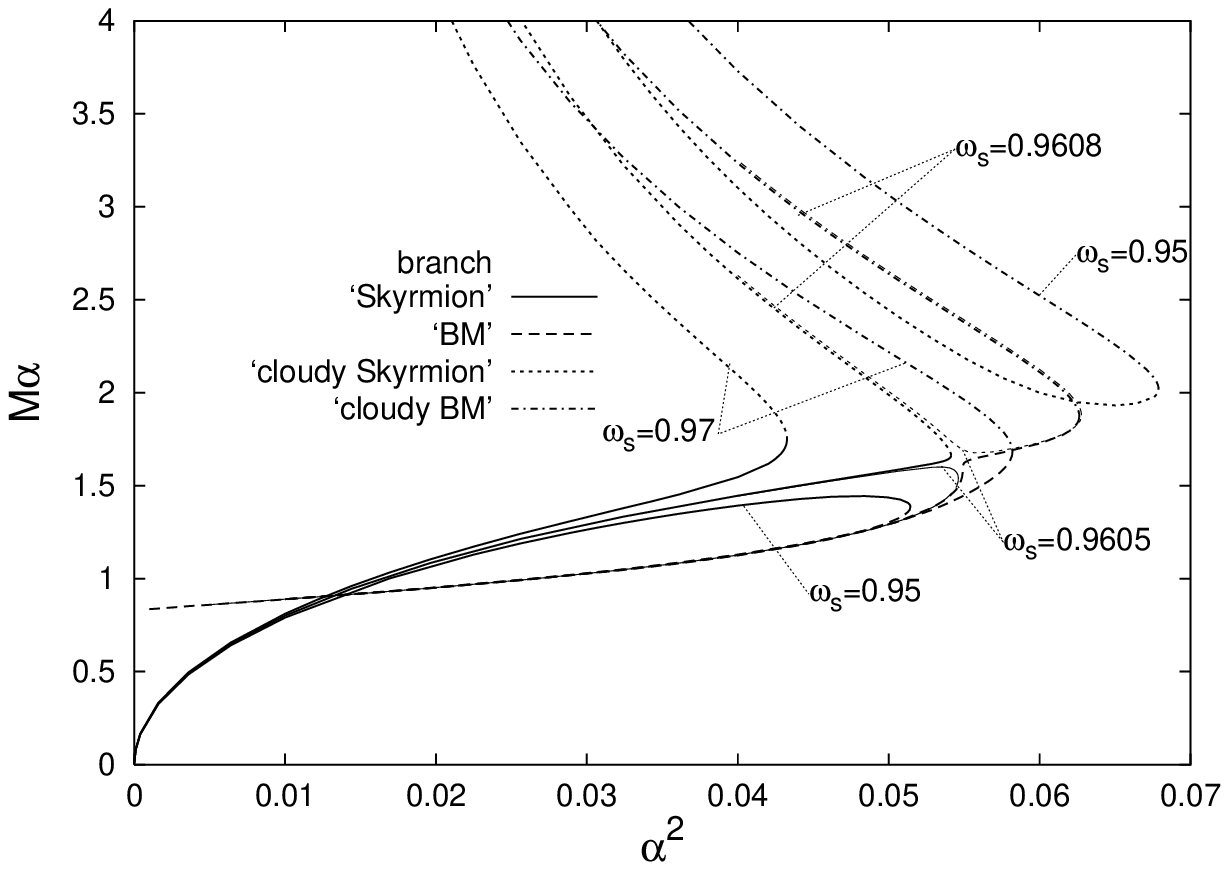}} \\
\epsfysize=5.0cm
(c) \mbox{\epsffile{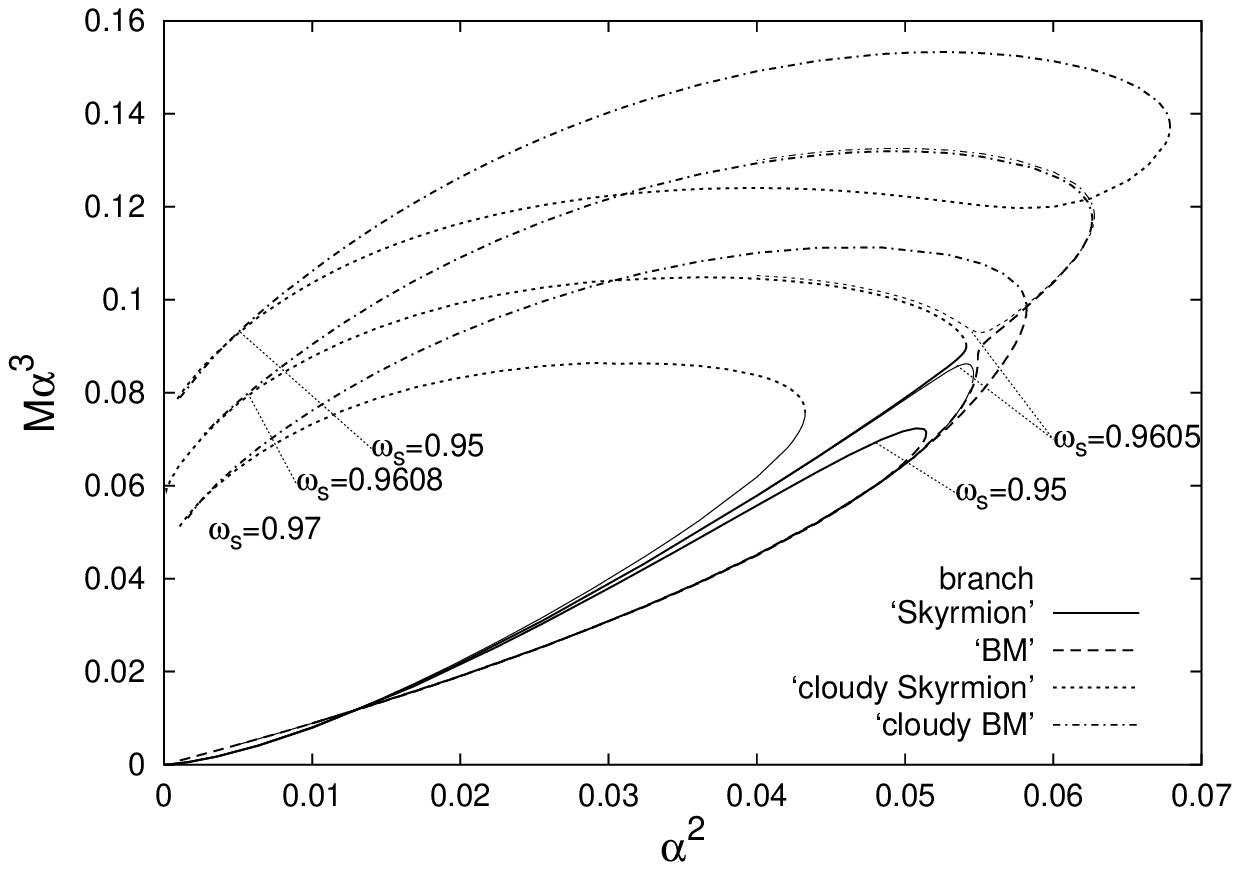}}
\epsfysize=5.0cm
(d) \mbox{\epsffile{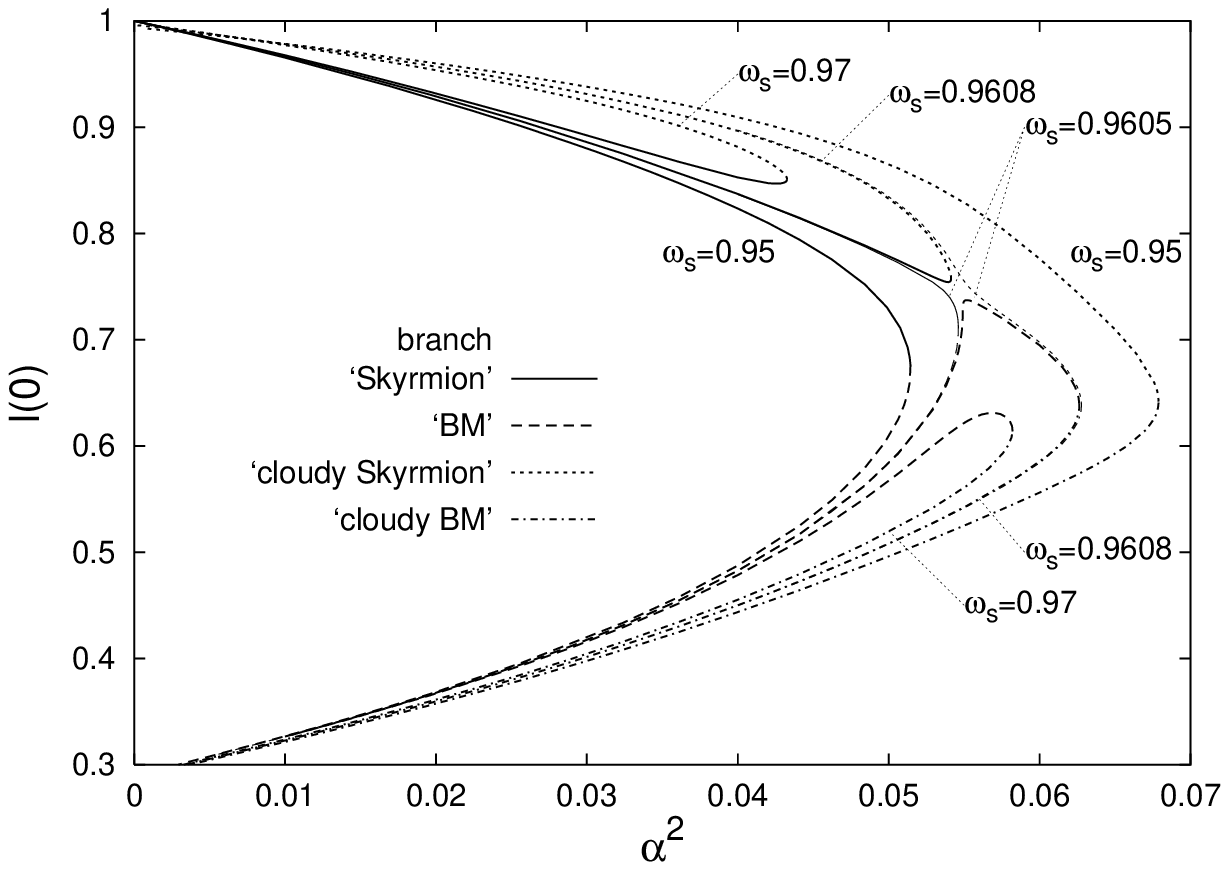}} \\
\caption{\label{Fig1}
The dimensionless mass $M$ (a),
the scaled masses $M\alpha$ (b) and $M\alpha^3$ (c),
and the  value of $l$ at the origin (d)
as function of $\alpha^2$ for several values of $\omega_s$.
}
\centering
\end{figure}

 Fig.~\ref{Fig1}(a) presents the mass $M$
for the `Skyrmion' branches (solid) which
merge either with the `BM' branches (dashed) or
the `cloudy Skyrmion' branches (dotted) when $\omega_s<\omega_s^{\rm cr}$
and  $\omega_s>\omega_s^{\rm cr}$, respectively.
Note that, as  $\alpha$ increases the mass decreases on the
`Skyrmion' branches, for small $\alpha$; but diverges
 on the `BM' and the `cloudy Skyrmion' branches as $\alpha$ tends to zero.
Also Fig.~\ref{Fig1}(a)  shows  the mass  of the `BM' branches
 merging with the `cloudy BM' branches (dash-dotted) when
$\omega_s>\omega_s^{\rm cr}$;
and   the mass of the `cloudy Skyrmion' branches  merging with the
`cloudy BM' branches when  $\omega_s<\omega_s^{\rm cr}$.

The scaled mass $M\alpha$ is plotted in  Fig.~\ref{Fig1}(b).
Note that, as $\alpha$ decreases along the `BM' branches
the scaled mass $M\alpha$ tends to a finite value which is equal to the mass
of the Bartnik-McKinnon solution.
Also, Fig.~\ref{Fig1}(b) shows the scaled mass $\alpha M$ of
the `BM' branches as it merges with the `cloudy BM' branches, and
of the `cloudy Skyrmion' branches as it merges with the `cloudy BM' branches.
Clearly $\alpha M$ diverges on the `cloudy Skyrmion' and the `cloudy BM'
branches for vanishing $\alpha$.

Fig.~\ref{Fig1}(c) reveals that in the limit $\alpha \to 0$ the
scaled mass $M\alpha^3$  of the `cloudy Skyrmion' and the `cloudy BM'
branch tends to a unique value which depends only on $\omega_s$.

Next we study  the quantity $l_0$ of Fig.~\ref{Fig1}(d) in order to
have a better understanding.
Following a solution along a `Skyrmion' branch which
merges with a `BM' branch,  $l_0$ decreases monotonically first
along the `Skyrmion' branch as $\alpha$ increases and then along
the `BM' branch as $\alpha$ decreases, to take the value
of the Bartnik-McKinnon solution as $\alpha \to 0$.
 In contrast, when a `Skyrmion' branch merges with a
`cloudy Skyrmion' branch, $l_0$ reaches a minimum on the `Skyrmion' branch
and increases on the `cloudy Skyrmion' branch as $\alpha$ decreases.
On the other hand, when the `BM' branch merges with a `cloudy BM' branch,
$l_0$ increases with increasing $\alpha$ along the `BM' branch
until it reaches a maximum and decreases with decreasing $\alpha$ along
the `cloudy BM' branch.
Finally, when a `cloudy Skyrmion' branch merges with a `cloudy BM' branch,
$l_0$ decreases monotonically if one follows the solutions first on the
`cloudy Skyrmion' branch with decreasing $\alpha$ and then
on the `cloudy BM' branch with increasing  $\alpha$.

Also, while we observe  that the scaled mass $M\alpha^3$ of the `cloudy Skyrmion'
and the `cloudy BM' branches tends to the same value as $\alpha$ tends
to zero --
this is  not true for the quantity $l_0$.
Thus, we conclude that the solutions approach different limits,
though with the same (scaled) mass.

\begin{figure}[h]
\centering
\epsfysize=5.0cm
(a) \mbox{\epsffile{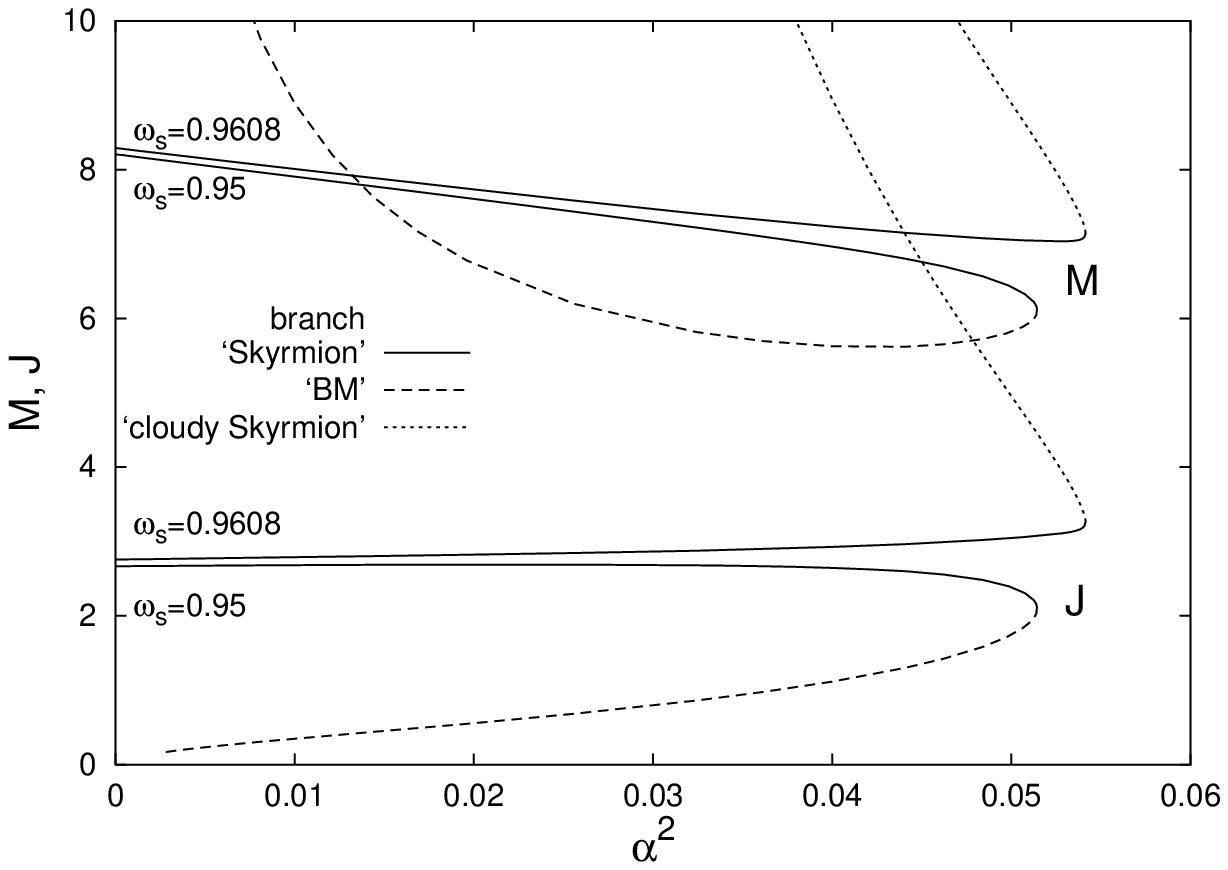}}
\epsfysize=5.0cm
(b) \mbox{\epsffile{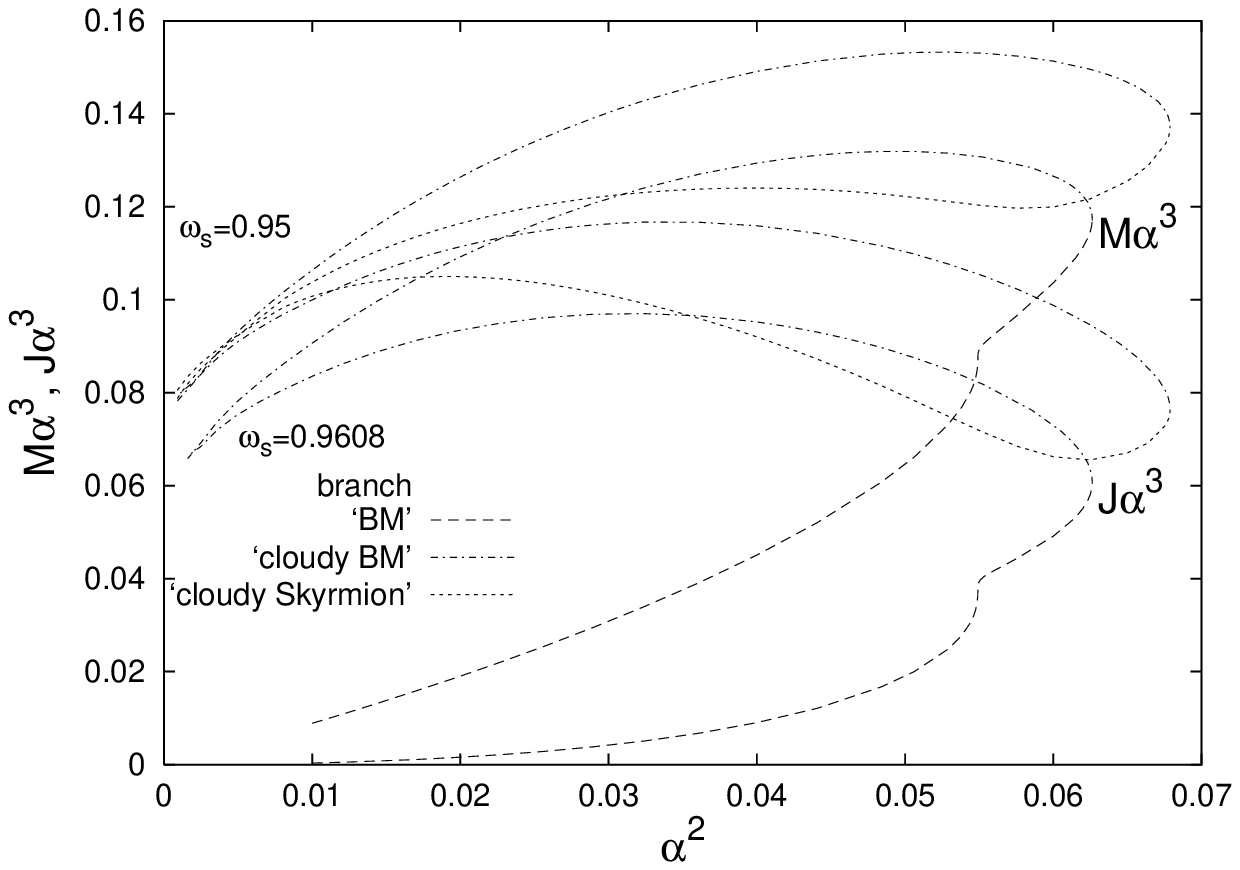}}
\caption{\label{Fig1a}
The dimensionless mass $M$ and angular momentum $J$ (a),
and the scaled mass $M\alpha^3$ and angular momentum $J\alpha^3$ (b)
 as function of $\alpha^2$ for $\omega_s=0.95$ and $\omega_s=0.9608$.
}
\centering
\end{figure}

In the following we compare the mass and the angular momentum.
We here restrict to $\omega_s=0.95<\omega_s^{\rm cr}$
and $\omega_s=0.9608>\omega_s^{\rm cr}$, as examples.
In particular, Fig.~\ref{Fig1a}(a) shows that
when $\omega_s=0.95 < \omega_s^{\rm cr}$ the
angular momentum decreases monotonically on the `Skyrmion' branch
as $\alpha$ increases and on the `BM' branch as $\alpha$ decreases,
while it tends to zero on the `BM' branch as $\alpha \to 0$.
In contrast, for $\omega_s=0.9608 > \omega_s^{\rm cr}$ the
angular momentum increases monotonically on the `Skyrmion' branch
as $\alpha$ increases and on the `cloudy Skyrmion' branch as
$\alpha$ decreases, while
in the limit $\alpha \to 0$ the angular momentum diverges
like $\alpha^3$.
Fig.~\ref{Fig1a}(b) presents the scaled mass
$M\alpha^3$ and angular momentum $J\alpha^3$ for the
`BM' and `cloudy BM' branches when $\omega_s=0.9608$ and
for the `cloudy Skyrmion' and `cloudy BM' branches when $\omega_s=0.95$.
As  $\alpha \to 0$
the (scaled) angular momentum tends to zero on the
`BM' branch, but takes finite values on the `cloudy' branches.

\begin{figure}[ht]
\centering
\epsfysize=4.5cm
(a) \mbox{\epsffile{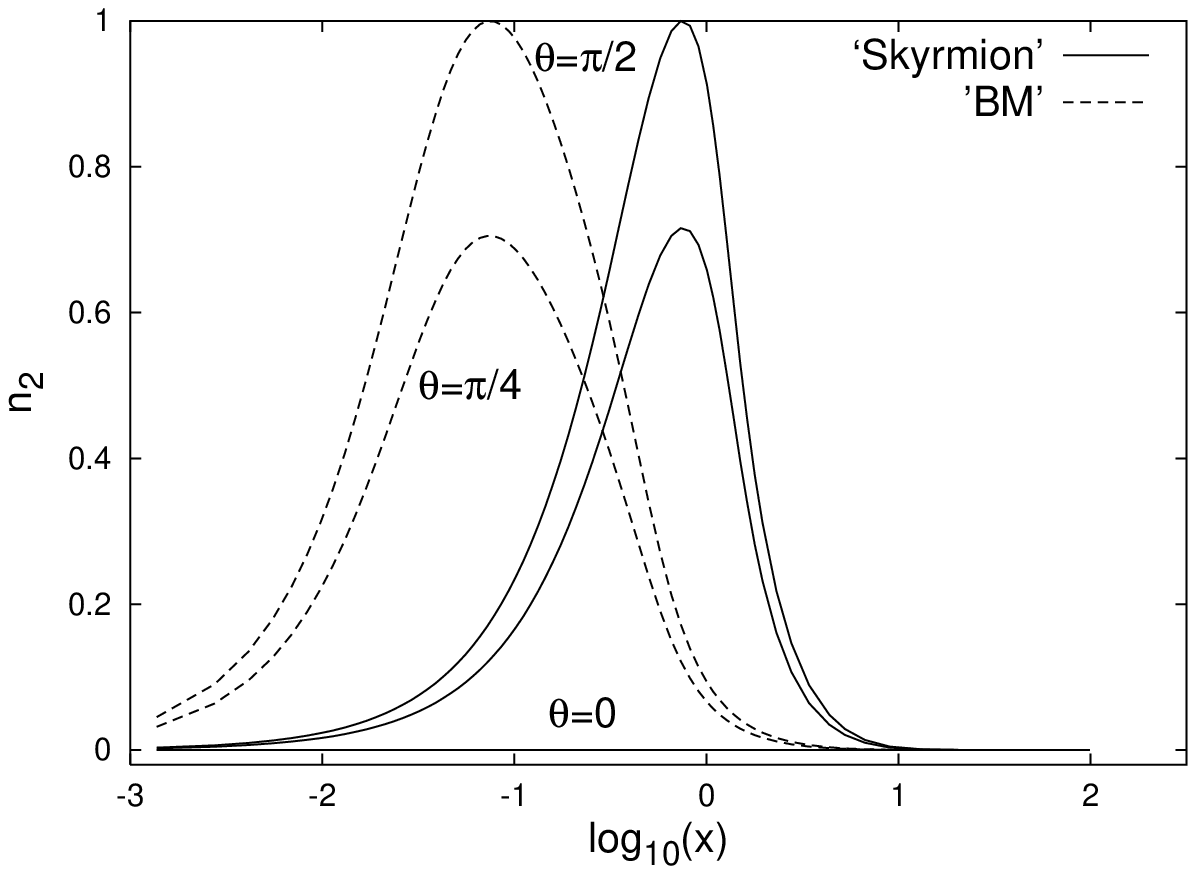}}
\epsfysize=4.5cm
(b) \mbox{\epsffile{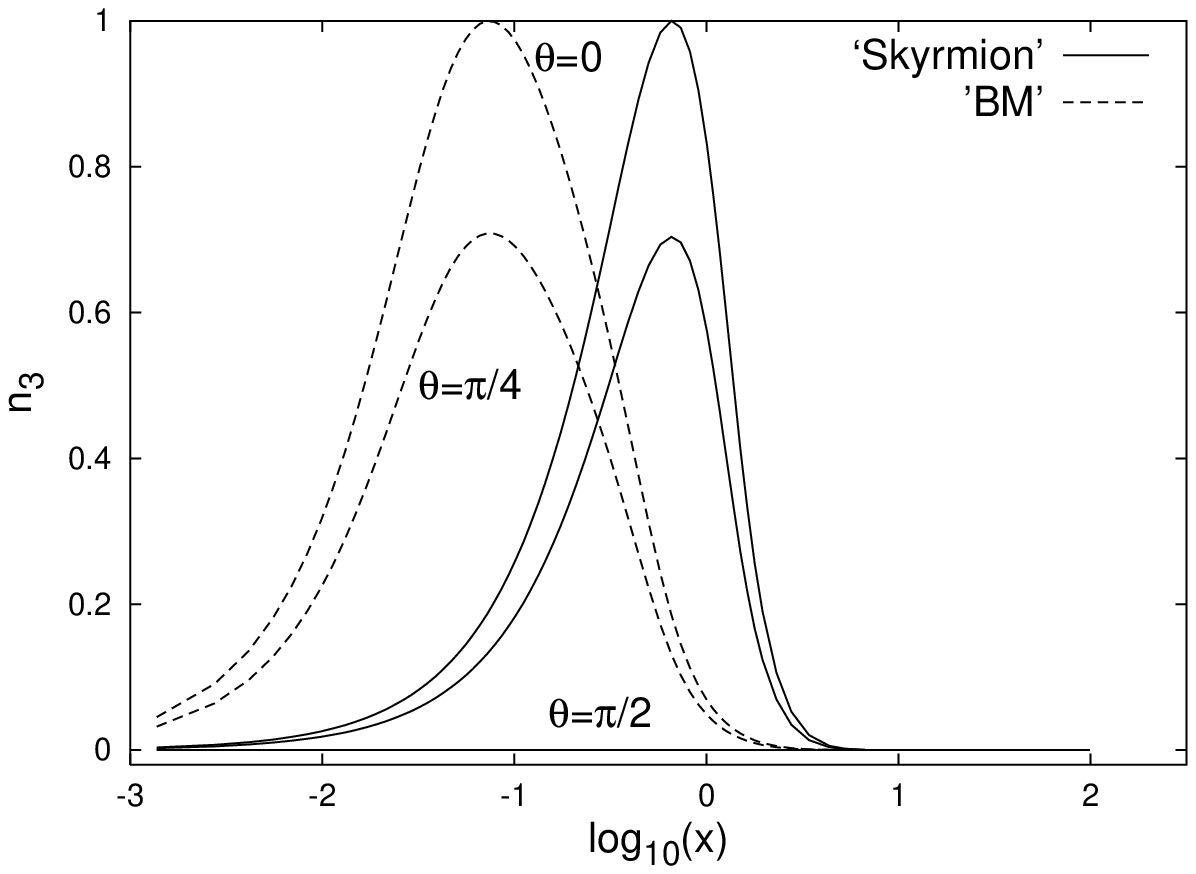}}\\
\epsfysize=4.5cm
(c) \mbox{\epsffile{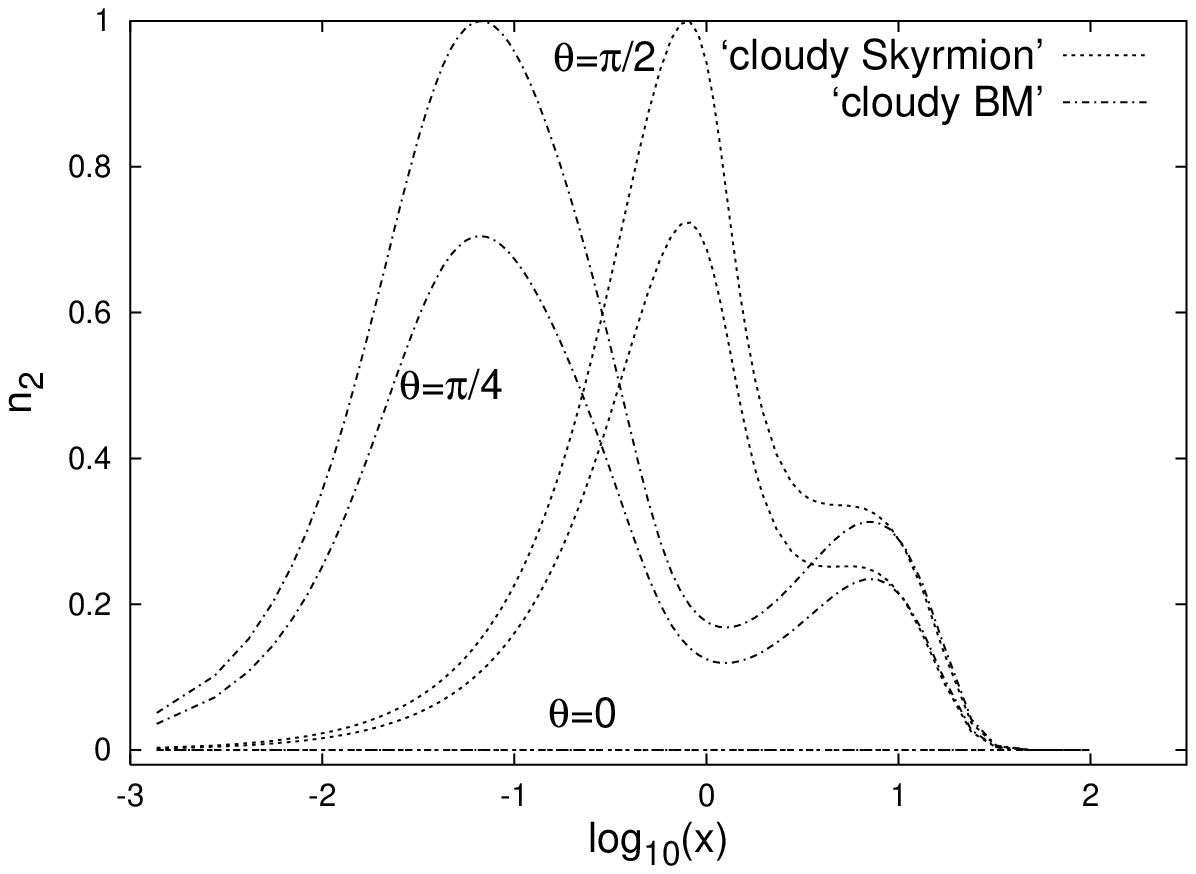}}
\epsfysize=4.5cm
(d)\mbox{\epsffile{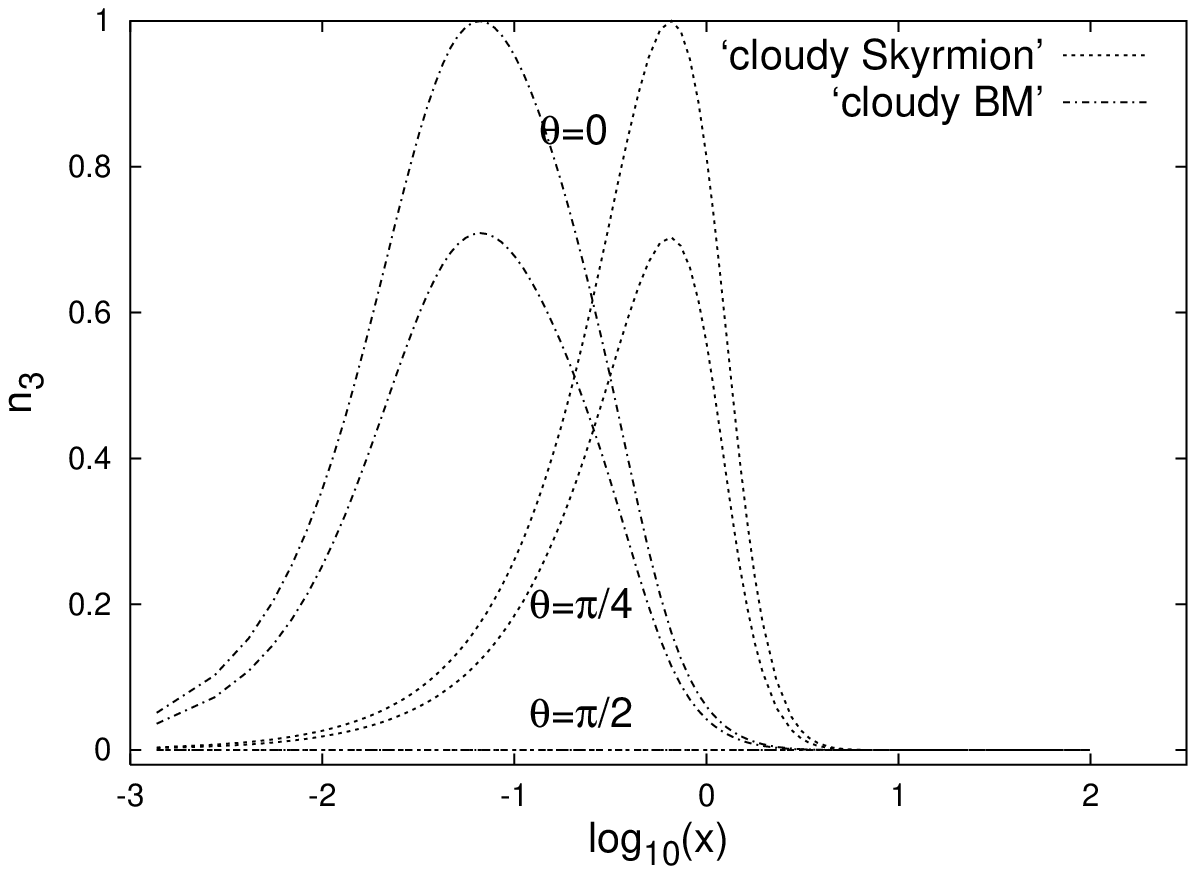}}
\caption{\label{Fig2}
The functions  $n_2$ (left) and $n_3$ (right) are plotted
for $\theta=0,\pi/4,\pi/2$ for the four solutions with
parameter values $\omega_s=0.95$ and $\alpha=0.1$.
}
\end{figure}
\begin{figure}[h!]
\centering
\epsfysize=4.5cm
(a) \mbox{\epsffile{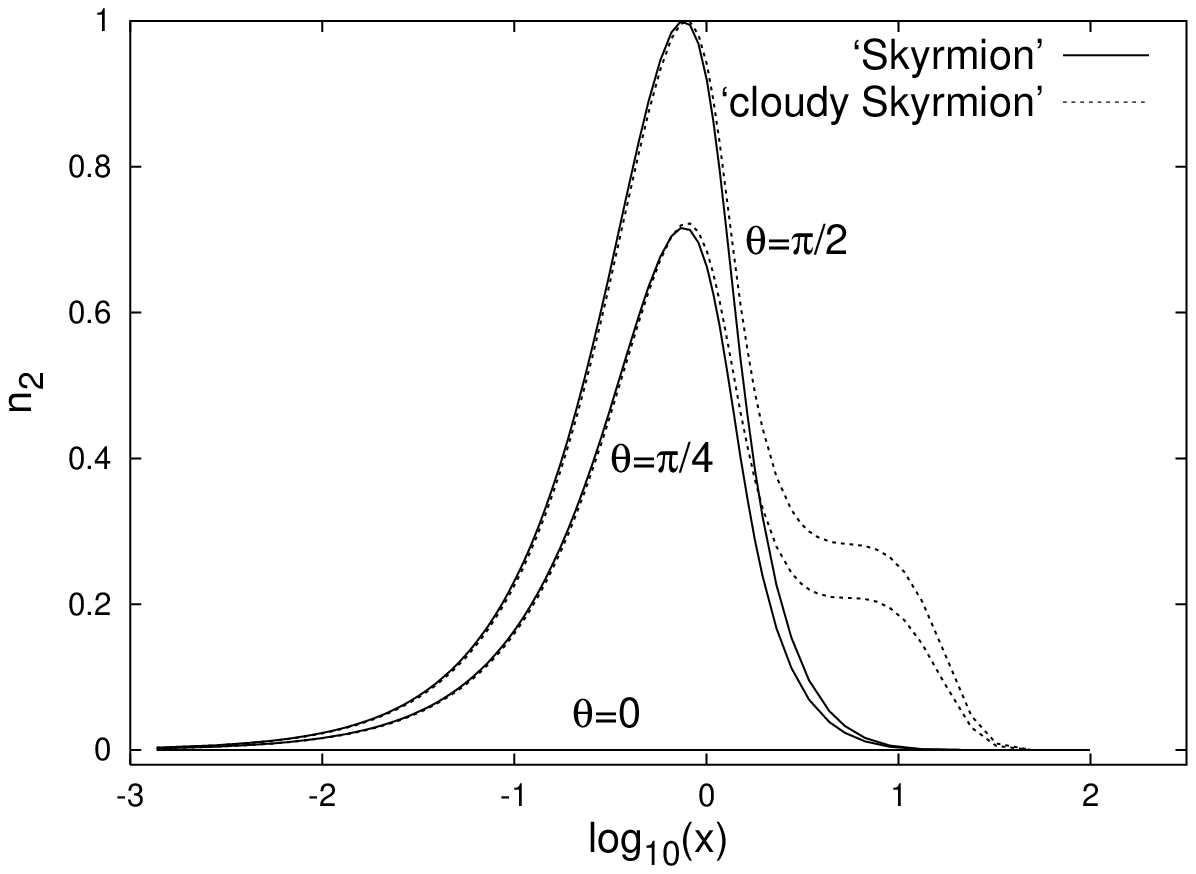}}
\epsfysize=4.5cm
(b) \mbox{\epsffile{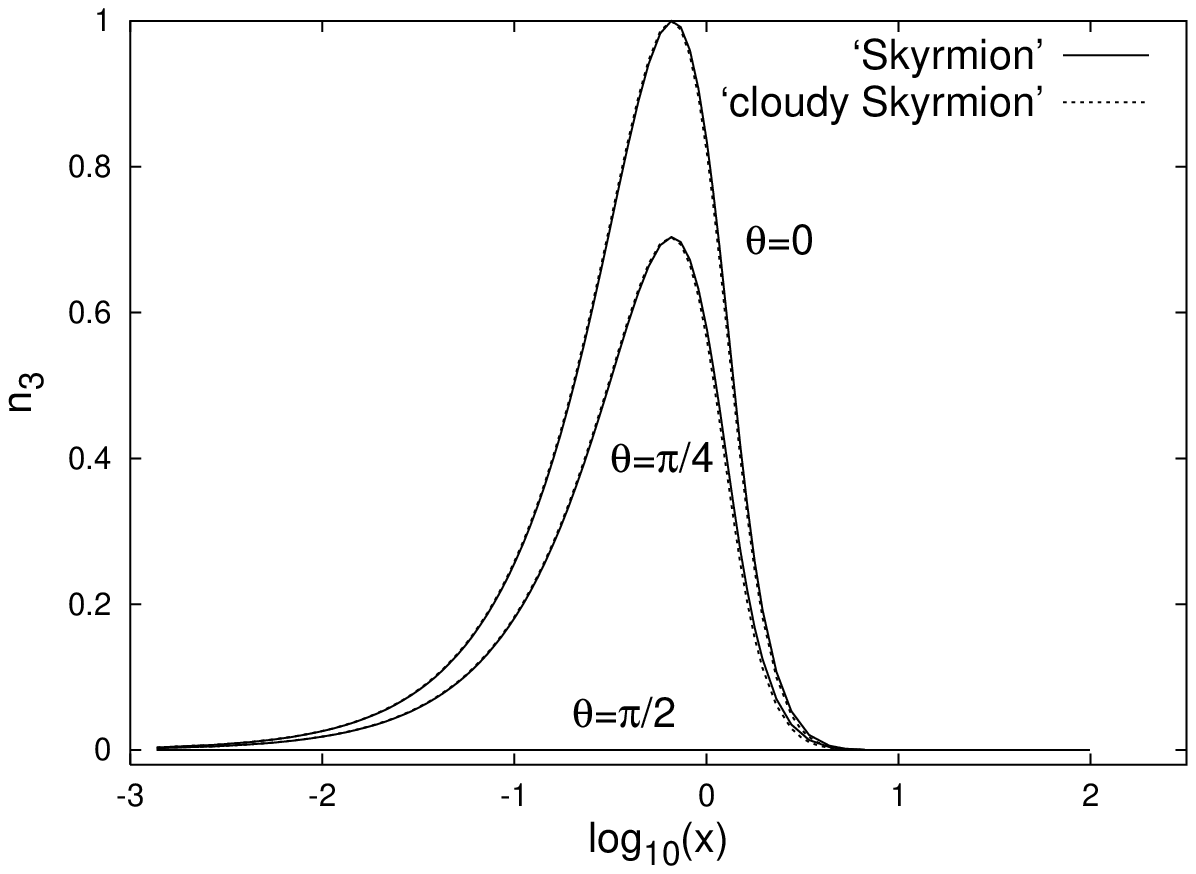}}\\
\epsfysize=4.5cm
(c) \mbox{\epsffile{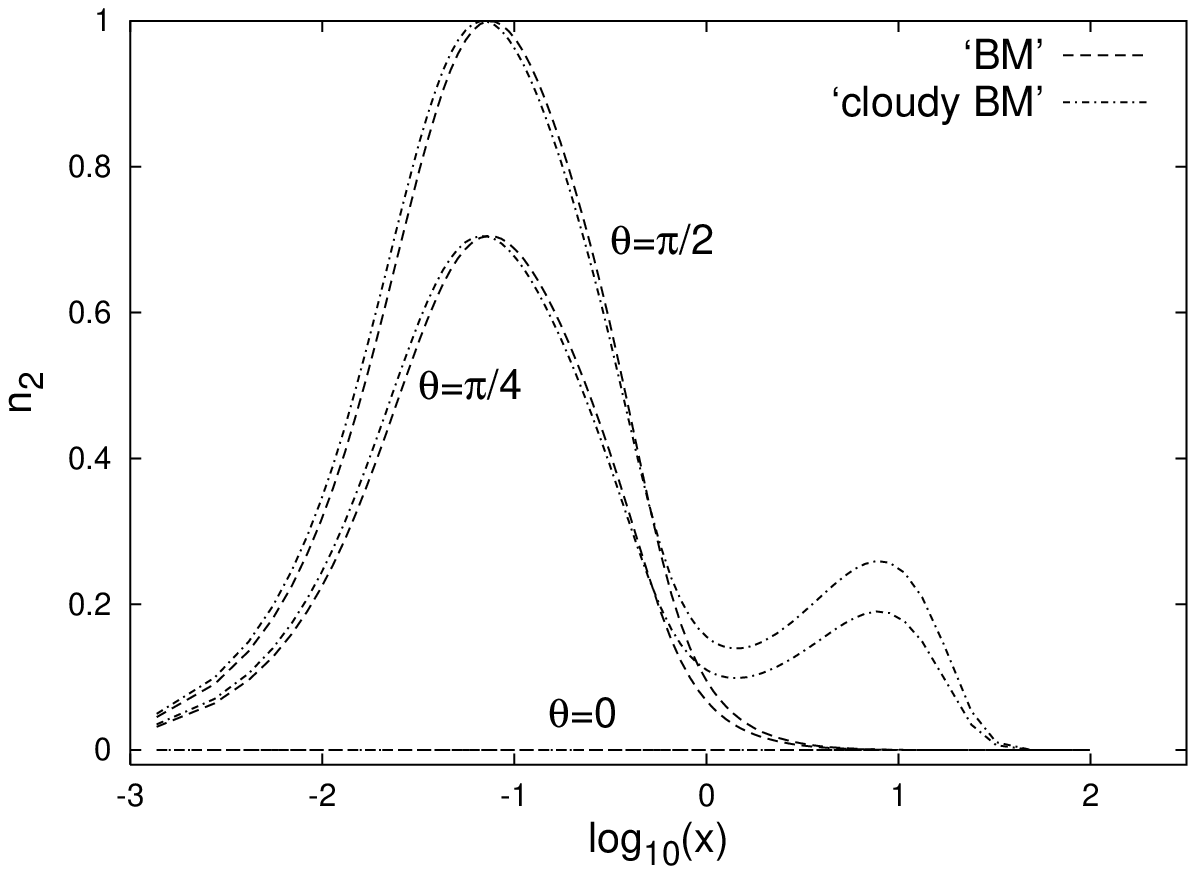}}
\epsfysize=4.5cm
(d) \mbox{\epsffile{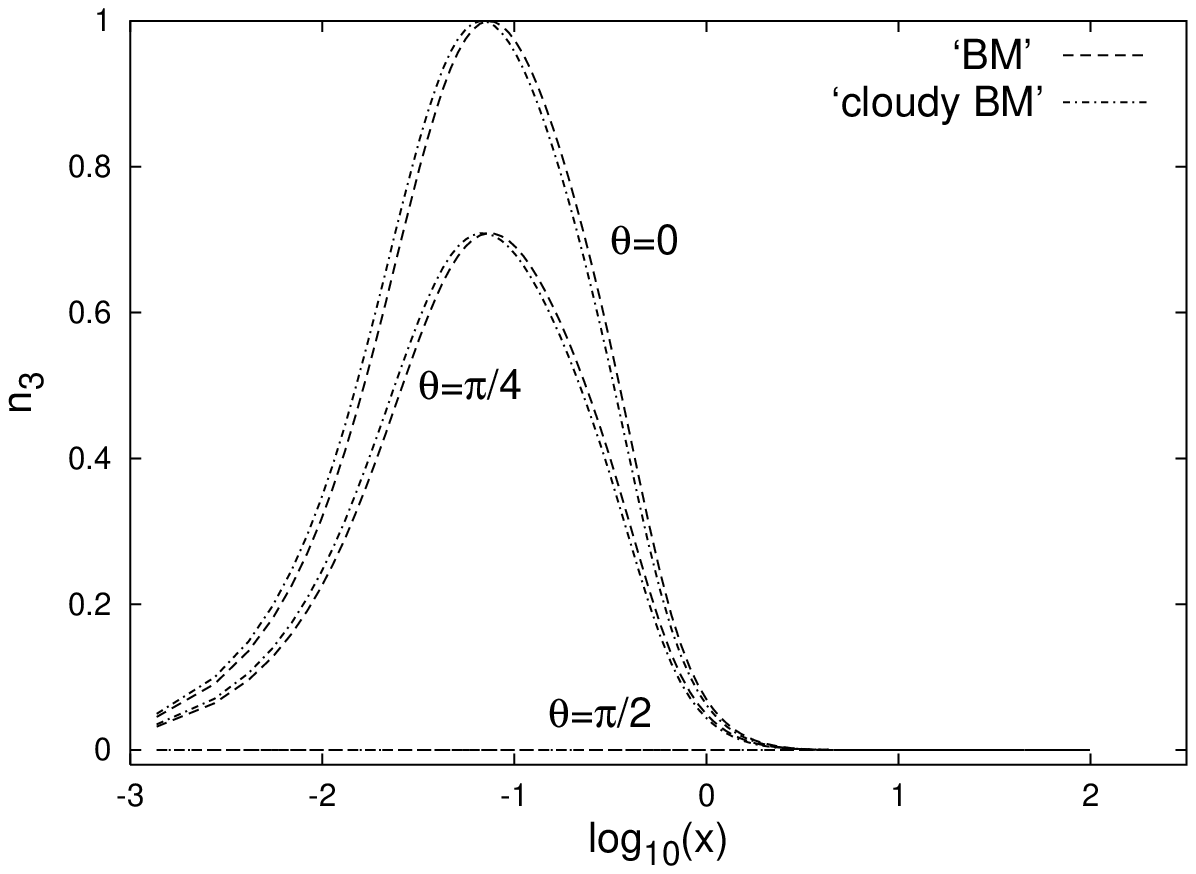}}
\caption{\label{Fig3}
Analogous as Figure \ref{Fig2} for  $\omega_s=0.9608$.
}
\end{figure}

The new interesting feature of the rotating gravitating Skyrmions
is the formation of  the `pion cloud'.
This can be  demonstrated by plotting  the  Skyrmion functions $n_2$ and $n_3$
when $\alpha=0.1$   for $\omega_s =0.95 < \omega_s^{\rm cr}$
and  $\omega_s =0.9608 > \omega_s^{\rm cr}$ as presented in
 Fig.~\ref{Fig2}  and Fig.~\ref{Fig3} (respectively).

In particular,   Figs.~\ref{Fig2}(a) and (b)  show that
on the `Skyrmion' branch the solution is close
to the flat space Skyrmion whereas on the `BM' branch it is close
to the (scaled) Bartnik-McKinnon solution.
In contrast,  the functions $n_2$ of the solutions on the `cloudy Skyrmion'
and `cloudy BM' branch  (presented in  Figs.~\ref{Fig2}(c))
 almost coincide at large $x$ where the
`pion cloud' forms.
However,  for small $x$ the functions
$n_2$ are similar in shape to the corresponding ones of
Fig.~\ref{Fig2}(a).
Finally, by comparing the functions $n_3$ of the `Skyrmion' and `BM'
solutions (of  Fig.~\ref{Fig2}(b)) with their `cloudy' counterparts
 (of  Fig.~\ref{Fig2}(d)) we observe that they are similar
in shape for all $x$ -- which means that  the  `pion cloud' does not
reflect itself in the function $n_2$.

Fig.~\ref{Fig3} presents the solutions on connected branches when
 $\omega_s > \omega_s^{\rm cr}$.
Note that, the function $n_2$ of both the `Skyrmion' and the
`cloudy Skyrmion' branch (plotted in Fig.~\ref{Fig3}(a)) almost coincide for
small values of $x$ and  differ for larger $x$,
where the `pion cloud' is apparent.
In contrast, the functions $n_3$ are almost identical for
 both solutions  as shown in Fig.~\ref{Fig3}(b).
Similar observations hold for the solutions of the `BM' branch and the
`cloudy BM'   branch  plotted in Fig.~\ref{Fig3}(c)-(d).

On the `cloudy BM' branch the scaled BM solution in the
core separates from the surrounding `pion cloud'  as $\alpha$ decreases.
Therefore, in what follows, we  show that  pure `pion cloud' solutions
 can be constructed numerically  by extracting the data of the `pion cloud'.
The ansatz for the `pion cloud' solution is given by
\begin{equation}
U =\cos(h) \identity +
     i \sin(h) \left( \cos(\vphi + \omega_s t) \tau_x +
                       \sin(\vphi + \omega_s t) \tau_y \right)  ,
\label{Umat_pi}
\end{equation}
where the boundary conditions for the profile function $h(x,\theta)$
follow from finite mass and regularity requirements and  read as
\begin{equation}
h(0,\theta)= 0 , \ \ \
h(\infty,\theta)= 0 , \ \ \
h(x,\theta=0)=0 , \ \ \
\partial_\theta h(x,\theta=\pi/2)=0 .
\end{equation}
For these solutions $n_3=0$ so  the
chiral matrix  can be regarded as a map from $S^3 \to S^2$ and thus,
 the `pion cloud' solutions have zero baryon number.
In addition, due to  the boundary conditions of the profile function
$h(x,\theta)$, the `pion cloud' solutions can be
deformed continuously to the vacuum.

\begin{figure}[h]
\centering
\epsfysize=7cm
\mbox{\epsffile{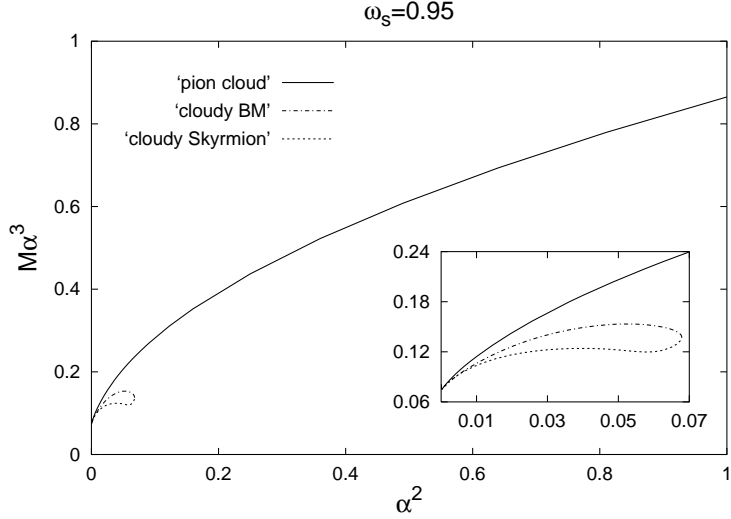}}
\caption{\label{Fig6}
The scaled mass $M\alpha^3$  for the `pion cloud' solutions
and the Skyrmion solutions on the `cloudy Skyrmion' and the `cloudy BM'
branches for $\omega_s=0.95$.
}
\centering
\end{figure}

Fig.~\ref{Fig6} presents the scaled mass $M\alpha^3$
of the `pion cloud' solutions and of the Skyrmions on the `cloudy' branches
 as function of $\alpha^2$ for fixed $\omega_s=0.95$.
Note that, for small $\alpha$ the masses of the `cloudy Skyrmion' and
`cloudy BM' solutions coincide with the mass of the `pion cloud' solution.
However, whereas the branches of the Skyrmion solutions exist only up to a
maximal value of $\alpha$, the `pion cloud' solutions exist
for arbitrarily large $\alpha$.

Indeed, as $\alpha$ increases the magnitude of $h$ decreases linearly
like $1/\alpha$.
So by expanding the Lagrangian up to quadratic order in $h$ yields
\begin{equation}
{\cal L}_h = \frac{R}{2 \alpha^2}-\frac{1}{2}\left[
\partial_x h \,\partial^x h +\frac{1}{x^2} \partial_\theta h \,
\partial^\theta h +\frac{f}{l x^2 \sin^2\theta} h^2
-\frac{1}{f} \left(\omega_s-\frac{\omega}{x}\right)^2 h^2 + \hat{m}_\pi^2 h^2 \right],
\label{lag_h}
\end{equation}
which is equivalent to the Lagrangian of the rotating boson star

\begin{equation}
{\cal L}_{BS} = \frac{R}{2 \alpha^2}-
\frac{1}{4}g^{\mu\nu}\left(
\pr_\mu \Phi^\ast \pr_\nu \Phi + \pr_\nu \Phi^\ast \pr_\mu \Phi\right)
-V(|\Phi|),
\label{lag_bs}
\end{equation}
for $\Phi = h(x,\theta)\, e^{i(\varphi - \omega_s t)}$ and
$V(|\Phi|)= |\Phi|^2  m_\pi^2/2$.
The limit $\alpha \to \infty$ of rotating boson stars has been studied
in Ref.~\cite{KKL} (though with different notation) where it was shown that the
 field equations become independent of the coupling parameter $\alpha$
after re-scaling $h=\hat{h}/\alpha$.
Moreover, it was argued in \cite{KKL} that
several branches of solutions exist in certain ranges of
$\omega_s$ which suggests that  several branches
of `pion cloud' solutions  might (also) exist, at least for large values of
$\alpha$.

In the limit $\alpha \to 0$ the scaled mass $M\alpha^3$ takes finite values.
By introducing the scaled radial coordinate $\xi = \alpha x$,
the Skyrme field equations reduces to the constraint
\begin{equation}
\sin^2(h)\left[\cos(h) \omega_s^2 -f  m_\pi^2\right]=0 \ ,
\label{constr_pc}
\end{equation}
as $\alpha \to 0$.
This implies that the Skyrme field function is either  $h=0$  or
 $\cos(h)=f  m_\pi^2/\omega_s^2$.
Note that none of these solutions can hold globally; the former
one yields the trivial solution and the latter is not consistent
with the asymptotic boundary conditions $h\to 0$ and $f \to 1$ for
$\omega_s^2 <  m_\pi^2$. However, both can hold locally. Indeed,
we find the solution
\begin{eqnarray}
\cos(h) & = & f  \frac{m_\pi^2}{\omega_s^2} \ , \ \ \ \ \
\{\xi, \theta\} \in {\cal D} = (0,\xi_0] \times (0, \pi )
\nonumber\\
h & = & 0  \ , \ \ \ \ \ \ \ \ \ \  \ \ \ \ \ {\rm elsewhere} \ ,
\nonumber
\end{eqnarray}
where $\xi_0$ depends on $\omega_s$.
Although this solution for $h$ is not continuous at the origin and on the
$z$-axis, the metric functions are continuous globally.
Moreover, in the domain ${\cal D}$ the function $h$ depends only on the
radial coordinate $\xi$. Consequently, the metric is spherically symmetric
and, for $\xi \geq \xi_0$, given by the Schwarzschild solution.

Thus, in the limit $\alpha \to 0$ the `pion cloud' becomes
confined to the finite domain ${\cal D}$. Outside of ${\cal D}$ the metric
is the vacuum solution determined by the connecting conditions at $\xi_0$.
To see the physical picture, however, we have to return to unscaled
coordinates. Then the domain  ${\cal D}$ extends over the whole space,
except the $z$-axis and infinity. Consequently, the `pion cloud'
occupies an increasing volume in space as $\alpha$ decreases.

\begin{figure}[h]
\centering
\mbox{(a)\epsfysize=5cm\epsffile{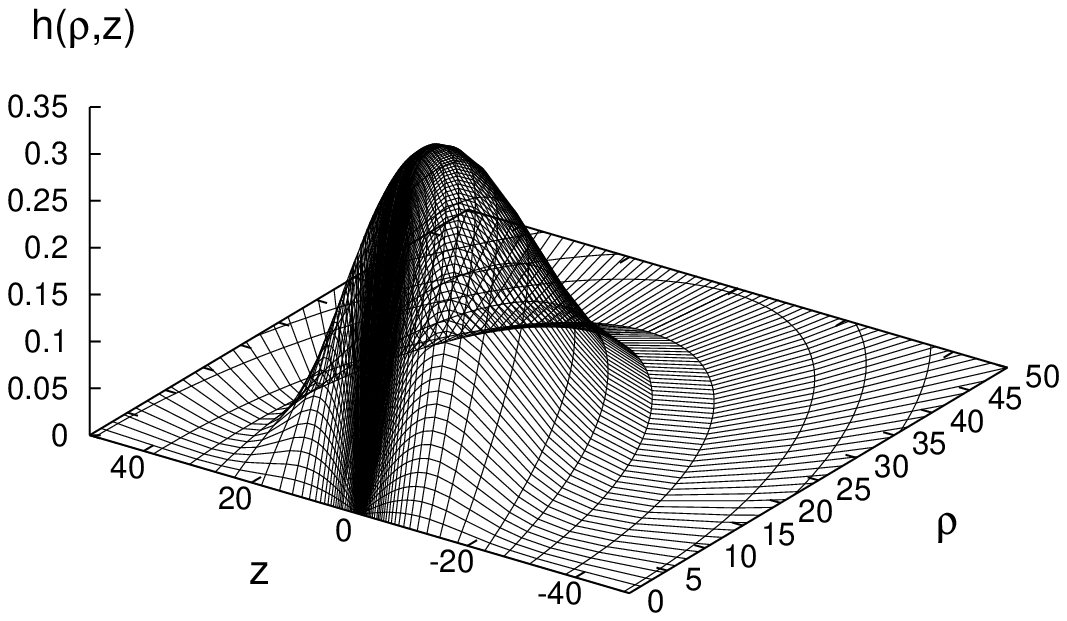}
(b)\epsfysize=5cm\epsffile{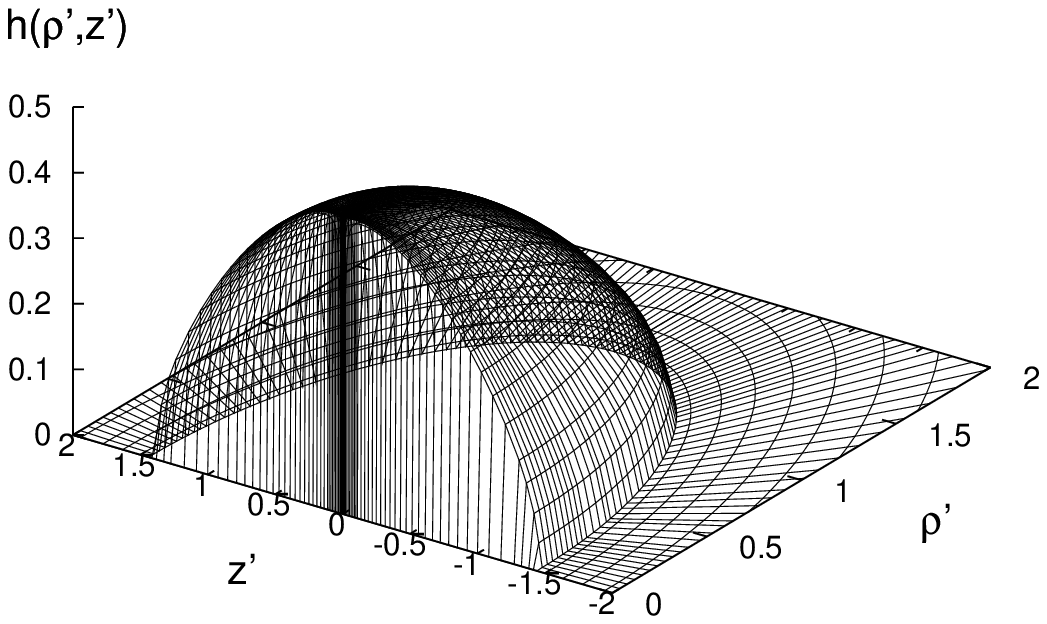}}
\caption{\label{Fig5}
(a) The `pion cloud' profile function $h(\rho,z)$  when
 $\omega_s=0.95$ and $\alpha=0.07$.
(b) Same as (a) for $\alpha=10^{-4}$ in scaled coordinates
$\rho' = \xi \sin\theta$, $z'=\xi\cos\theta$.
}
\centering
\end{figure}

The profile function $h$ is plotted in Fig.~\ref{Fig5}(a) as function of
the  cylindrical coordinates $\rho = x \sin\theta$, $z=x\cos\theta$
for $\omega_s=0.95$ and $\alpha=0.07$ and in Fig.~\ref{Fig5}(b) as function of
the scaled coordinates $\rho' = \xi \sin\theta$, $z'=\xi\cos\theta$
for such parameter values that the constraint is almost satisfied
(ie. $\omega_s=0.95$ and $\alpha=10^{-4}$).

Finally we state that the `pion cloud' solutions do not exist
for arbitrarily small $\omega_s$.
In fact, we observed that the
coefficient of the second order derivative term in the Skyrme field
equation
\begin{equation}
\left[f^2 \sin^2(h) -\left(
\left(\omega_s-\frac{\omega}{r}\right)^2\sin^2(h)-f \right)
r^2 l \sin^2\theta \right]
\left(\partial^2_{rr} h +\frac{1}{r^2} \partial^2_{\theta\theta} h \right)
+ \cdots = 0
\end{equation}
develops a zero at some point on the $\theta=\pi/2$ axis, when
$\omega_s$  decreases to a critical value, and no solution exists
for $\omega_s$ below the critical value. Therefore, the `pion
cloud' solutions do not possess a static limit.

\section{Conclusions}

We studied stationary rotating solutions of the Einstein-Skyrme theory.
These solutions are asymptotically flat, globally regular and axially symmetric.
Branches of stationary rotating Skyrmions emerge from corresponding branches
of static Skyrmions, when the rotational frequency is increased form zero.
If the rotational parameter $\omega_s$ is smaller than a critical value,
the rotating Skyrmions on these branches behave similar to their static counterparts,
when the coupling to gravity is varied. However, additional branches of solutions
exist, which do not have a flat space limit. These branches are characterised by the
formation of a `pion cloud' for small  values of the coupling $\alpha$.

In addition to the rotating Skyrmions with baryon number one,
we found new solutions in the topologically trivial sector.
These `pion cloud' solutions are also asymptotically flat and globally regular,
but possess neither a flat limit nor a static limit.
In contrast to the Skyrmions the  `pion cloud' solutions exist
for arbitrary coupling, ie. $0 <\alpha < \infty$.

Axially symmetric rotating Skyrmions with higher baryon number $B>1$
should easily be obtained by the replacement $\varphi \to B\varphi$
in the ansatz Eq.~(\ref{Umat}).

Einstein-Skyrme theory also possesses black hole solutions.
So far the Skyrmion black holes have been studied in the static limit only.
Families of black hole solutions emerge from the globally static regular Skyrmions,
when the horizon radius is increased from zero.
Similarly, we expect several families of stationary rotating Skrymion black holes
to emerge from the globally regular rotating Skyrmions on the different branches
obtained here \cite{IKKbh}.
Moreover, one may speculate that also stationary rotating black holes
emerge from the `pion cloud' solutions.

{\bf Acknowledgment}

BK gratefully acknowledges support by the DFG under contract
KU612/9-1 and TI thanks Oldenburg University for its hospitality.


\begin{thebibliography}{99}
\bibitem{overview}
for an overview see e.g.
 M.~S.~Volkov and D.~V.~Gal'tsov,
  Phys.\ Rept.\  {\bf 319}, 1 (1999).

\bibitem{rotbh}
 M.~S.~Volkov and N.~Straumann,
  Phys.\ Rev.\ Lett.\  {\bf 79} (1997) 1428;\\
 O.~Brodbeck, M.~Heusler, N.~Straumann and M.~S.~Volkov,
  Phys.\ Rev.\ Lett.\  {\bf 79} (1997) 4310;\\
 B.~Kleihaus and J.~Kunz,
  Phys.\ Rev.\ Lett.\  {\bf 86} (2001) 3704;\\
 B.~Kleihaus, J.~Kunz and F.~Navarro-Lerida,
  Phys.\ Lett.\ B {\bf 599} (2004) 294;\\
 B.~Kleihaus, J.~Kunz and F.~Navarro-Lerida,
  Phys.\ Rev.\ D {\bf 69} (2004) 064028;\\
 B.~Kleihaus, J.~Kunz and F.~Navarro-Lerida,
  Phys.\ Rev.\ Lett.\  {\bf 90} (2003) 171101.

\bibitem{vdBijRadu}
 J.~J.~Van der Bij and E.~Radu,
  Int.\ J.\ Mod.\ Phys.\ A {\bf 17} (2002) 1477.
\bibitem{VolWoh}
 M.~S.~Volkov and E.~W\"ohnert,
  Phys.\ Rev.\ D {\bf 67} (2003) 105006.
\bibitem{vdBijRadu2}
 J.~J.~Van der Bij and E.~Radu,
  Int.\ J.\ Mod.\ Phys.\ A {\bf 18} (2003) 2379.

\bibitem{PRT}
 V.~Paturyan, E.~Radu and D.~H.~Tchrakian,
  Phys.\ Lett.\ B {\bf 609} (2005) 360.
\bibitem{KKN}
 B.~Kleihaus, J.~Kunz and U.~Neemann,
  Phys.\ Lett.\ B {\bf 623} (2005) 171.

\bibitem{BKS}
R.A. Battye, S. Krusch and P.M. Sutcliffe, Phys. Lett. B {\bf 626} (2005) 120.
\bibitem{RaTch}
 E.~Radu and D.~H.~Tchrakian,
  Phys.\ Lett.\ B {\bf 632} (2006) 109

\bibitem{spinSk}
 E.~Braaten and J.~P.~Ralston,
  Phys.\ Rev.\ D {\bf 31} (1985) 598;\\
  M.~Bander and F.~Hayot,
  Phys.\ Rev.\ D {\bf 30} (1984) 1837;\\
   K.~F.~Liu and J.~S.~Zhang,
  Phys.\ Rev.\ D {\bf 30} (1984) 2015.

\bibitem{gravSk}
 H. Luckock and I. Moss,
 Phys. Lett. {\bf B176} (1986) 341;\\
 H. Luckock,
 {\it Black hole skyrmions},
 Proceedings of the 1986 Paris-Meudon Colloquium,
 eds. H.~J. de Vega, and N. Sanchez,
 (World Scientific, Singapore, 1987);\\
 S. Droz, M. Heusler and N. Straumann,
 Phys. Lett. {\bf B268} (1991) 371.

\bibitem{Heussler}
 M. Heusler, S. Droz and N. Straumann,
 Phys. Lett. {\bf B271} (1991) 61.

\bibitem{Bizon}
 P. Bizon and T. Chmaj,
 Phys. Lett. {\bf B297} (1992) 55.

\bibitem{BM}
 R.~Bartnik and J.~McKinnon,
  Phys. Rev. Lett. {\bf 61} (1988) 141.

\bibitem{fidisol}
 W. Sch\"onauer and R. Wei\ss,
 J. Comput. Appl. Math {\bf 27} (1989) 279;\\
 M. Schauder, R. Wei\ss \ and W. Sch\"onauer,
 The CADSOL Program Package, Universit\"at Karlsruhe,
 Interner Bericht Nr. 46/92 (1992).
\bibitem{KKL}
  B.~Kleihaus, J.~Kunz and M.~List,
  Phys.\ Rev.\ D {\bf 72} (2005) 064002.

\bibitem{IKKbh}
 T.~Ioannidou, B.~Kleihaus and J.~Kunz,
in preparation.

\end{thebibliography}
\end{document}